\documentclass{article}
\usepackage{amsfonts,amssymb,amsmath}
\usepackage[dvips]{epsfig}
\usepackage[T1]{fontenc}
\usepackage[latin1]{inputenc}
\usepackage{graphicx}
\usepackage[english]{babel}
\usepackage{amsmath}
\usepackage{amssymb}
\usepackage{amsfonts}
\usepackage{color}

\begin{document}

\title{\bf Emergent Universe in the Braneworld Scenario }
\author{Y. Heydarzade$^{1}$\thanks{email: heydarzade@azaruniv.edu}~, H. Hadi$^{1}$,
~F. Darabi$^{1,3}$\thanks{email: f.darabi@azaruniv.edu; (Corresponding author)}  and A. Sheykhi$^{2}$\thanks{email: asheykhi@shirazu.ac.ir}
\\{\small $^1$Department of Physics, Azarbaijan Shahid Madani University, Tabriz, Iran}\\ {\small $^2$Physics Department and Biruni Observatory, College of Sciences, Shiraz University, Shiraz 71454, Iran}\\ {\small $^3$Research Institute for Astronomy and Astrophysics of Maragha (RIAAM), Maragha 55134-441, Iran}}
\date{\today}

\maketitle
\begin{abstract}
According to Padmanabhan's proposal, the difference between the
surface degrees of freedom and the bulk degrees of freedom in a
region of space may result in the acceleration of Universe
expansion through the relation $\Delta V/\Delta t = N_{\rm sur}-N_{\rm
bulk}$  where $N_{\rm
bulk}$ and  $N_{\rm sur}$ are  referred to the degrees of freedom related
to the matter and energy content inside the bulk and surface area, respectively \cite{Pad1}. In this paper, we study the
dynamical effect of the extrinsic geometrical embedding of an
arbitrary four dimensional brane in a higher dimensional bulk
space and investigate the corresponding degrees of freedom. Considering
the modification of Friedmann equations arising from a general
braneworld scenario, we obtain a correction term in Padmanabhan's
relation,
denoting the number of degrees of freedom related to the extrinsic
geometry of the brane embedded in higher dimensional spacetime as
$\Delta V /\Delta t=N_{\rm sur}-N_{\rm bulk}-N_{\rm extr}$ where $N_{\rm extr}$ is referred to the degree of freedom related
to the extrinsic geometry of the brane while $N_{\rm sur}$ and $N_{\rm bulk}$
are as well as before.
Finally, we study the validity of the first and second laws of
thermodynamics for this general braneworld scenario in the state of
thermal equilibrium and in the presence of confined matter fields
to the brane with the induced geometric matter fields.
\\
Pacs: 98.80.Cq, 98.80.-k.
\end{abstract}

\section{Introduction}

Recent researches support the idea that the gravitational field equations
can be derived in the same way that the equations of an emergent phenomena like
fluid mechanics or elasticity are obtained \cite{14, 15}. 
According to emergent gravity paradigm the gravitational field equations can be derived
from thermodynamic principle \cite{T1, T2}. In this way, Padmanabhan has treated the  Einstein field equations as  emergent, while the existence
of a spacetime manifold, its metric and  curvature have been assumed \cite{16}.
In a cosmological context, it has been argued that the accelerated expansion of the Universe
can be derived from the difference between the surface  and 
bulk degrees of freedom through the relation
$\Delta V/\Delta t = N_{\rm sur}-N_{\rm
bulk}$, in which $N_{\rm
bulk}$ and   $N_{\rm sur}$ are  referred to the degrees of freedom related
to matter and energy content (or dark matter and dark energy) inside the bulk and surface area, respectively \cite{Pad1}. In order
to explain the present accelerated expansion of the Universe, which is in agreement with different data set or observational data \cite{21}, different
models have been proposed.
One of these models\ is the dark energy model which admits that the
universe is dominated by a dark fluid with negative pressure. However, there are several dark energy models such as dynamical dark energy \cite{DDE},
quintessence \cite{Q} and k-essence \cite{K}, for a review the reader is referred to \cite{review}. Also, the LCDM model or the
concordance model is a particular case of dark energy that is
parameterized by a cosmological constant $\Lambda$ with the equation of state
parameter equal minus one, i.e $p=-\rho$. The strong energy condition, i.e $\rho+3p>0$, is
violated by the dark energy because of
demanding for the positive acceleration of the Universe through the second Friedmann equation $\frac{\ddot
a}{a}= -\frac{4\pi G}{3}(\rho+ 3p)$. Another approach
lies in the framework of modified gravity theories which describe the present acceleration of the Universe such as $f(R)$ gravity 
\cite{f(R)}, $f(T)$ gravity  
\cite{f(T)}, Ho\v{r}ava-Lifshitz gravity  
\cite{HL},   Gauss-Bonnet gravity
\cite{GB},  Weyl gravity
\cite{W},  Lovelock gravity
\cite{LG},  massive gravity \cite{MG}   
and braneworld scenarios \cite{braneworld11}. In these models,
 there is no need for the introduction of an ad-hoc component usually called as 
dark energy with unusual features.  In these models, some additional terms are considered in the gravitational Lagrangian which lead to the modification
of the gravitational
theory resulting in an effective dark energy sector with a geometrical origin.
 One can also find such models in the low energy limit of heterotic string theory \cite{23}. All of these models admit a series of conditions coming
from various laws of physics such as thermodynamics laws \cite{24} or astrophysical
data.

 To explain  the structure of spacetime
and its relation with thermodynamics of the system,
 one
can refer to four laws of black hole mechanics which are derived from the classical Einstein
field equations. These four laws are  analogous to those of
thermodynamics \cite{17}.
Discovery of the quantum Hawking radiation \cite{18} turns out that this analogy is an identity. By
 deriving the Einstein field equation from the relation
of entropy and horizon area together with the thermodynamic law of
$Q= TdS$ which connects the heat $Q$, the entropy $S$, and the temperature $ T$,  Jacobson showed that the classical general relativity is a kind of thermodynamics where
the surface gravity is a temperature \cite{19}.
The generalized second law of thermodynamics is specially investigated in different modified gravity models. For example,
we can refer to the investigations  devoted to the study of generalized second law (GSL) of thermodynamics in $f(T)$ gravity models in which two types
of horizons, are used to check the validity of the generalized second law
of thermodynamics  with corrected entropies \cite{20}. One can also find
that in the state of thermal equilibrium, in Kaluza- Klein universe which is composed of dark matter and dark energy, the validity of the laws of thermodynamics are true \cite{sharif}. The
investigations on the deep connection between gravity and thermodynamics have been widely considered in the  cosmological context where it has been
shown that in the form of the first law of thermodynamics on apparent horizon, the differential
form of the Friedmann equation in the FRW universe can be written  \cite{25,
26, 27, 28, 29, 30, 31, 32, 33}.
 The
GSL in an accelerating universe related to the apparent horizon has been considered
in \cite{34,35,36}. It was discussed in \cite{35},  that in contrast to the case of the apparent horizon,
the general second law  of thermodynamics breaks down in the case of a universe  enveloped by the event horizon with the usual definitions of entropy and temperature. This study reveals that from the thermodynamical point of view,  in an accelerating universe with spatial curvature, the apparent horizon is a
physical boundary. Also, the general expression
of temperature at apparent horizon of FRW universe, allows one to show that
 the GSL holds in Einstein, Gauss-Bonnet and more general Lovelock gravity \cite{37}. Also, the GSL of thermodynamics   in the framework of braneworld scenarios  is studied in \cite {38}. One can find other studies
on the GSL of thermodynamics  in \cite{39,40,41}.

In particle physics, the warped product geometries, well known as Randall-Sundrum models, are very important \cite{RS}.  In these models, it is imagined that our real world is a higher-dimensional universe described by a warped geometry with $Z_{2}$ symmetry. The standard gauge interactions are confined to the four dimensional brane embedded in a higher dimensional bulk space where gravitons are propagating through the extra dimensions.
More specifically, our universe is assumed to be a five-dimensional anti-de Sitter space where the elementary particles, except for the gravitons, are localized on a $(3 +1)$-dimensional brane or branes.

In this paper, we consider a general braneworld model which provides a geometrical origin for dark energy or accelerating expansion of the Universe \cite{Maia1}.
Considering the modification of Friedmann equations resulted from this general
braneworld scenario, we obtain a correction term on Padmanabhan's
relation. This paper is organized as follows: In section 2, we introduced
general geometrical setup of the braneworld. In section
3, this braneworld model is studied under Israeal-Darmois-Lanczos junction condition, which provides  the $Z_2$ symmetry, and the corresponding number of degrees of freedom related to the extrinsic
geometry of such a brane model is obtained. In section 4,
we find the correction term to the Padmanabhan's relation in our general braneworld model which does not  have any specific junction condition.
In section 5, we explored the thermodynamics of such a general brane model.
At last, in section 6,  we presented our concluding remarks.

 \section{General Geometrical Setup of the Braneworld}

The effective Einstein-Hilbert action for the $4D$ spacetime
$(\mathcal{M}_{4},g)$ embedded in an $nD$ bulk space
$(\mathcal{M}_{n},\mathcal{G})$ can be written
\begin{equation}\label{act}
I_{EH}=\frac{1}{2\alpha_{*}}\int
d^{n}x\sqrt{-\mathcal{G}}\mathcal{R}
+\int_{\Sigma} d^{4}x\sqrt{-g}\mathcal{L}_{m},
\end{equation}
where $\alpha_{*}$, $\mathcal{R}$ and
$\mathcal{L}_m$ are respectively as gravitational coupling constant in the bulk space,
bulk Ricci scalar and the
Lagrangian density of the matter fields confined to the brane.
Variation of the action
(\ref{act}) with respect to the bulk metric
$\mathcal{G}_{AB} (A,B=0,...,n-1)$ leads to the following field
equations for the bulk space
\begin{equation}
G_{AB}=\alpha_{*}S_{AB,}
\end{equation}
where $S_{AB}$ is
\begin{equation}
S_{AB}=T_{AB}+\frac{1}{2}\mathcal{V}\mathcal{G}_{AB},
\end{equation}
and $T_{AB} = -2 \frac{\delta \mathcal{L}_m}
{\delta g_{AB}} + g_{AB}\mathcal{L}_m$ is the energy-momentum tensor of the matter
fields confined to the four dimensional brane through the action of the confining potential $\mathcal{V}$. The confining potential $\mathcal{V}$ satisfies three
general conditions: (I) It has a deep minimum on the original non-perturbed 
brane (we will discuss  on the original non-perturbed and perturbed geometry
in the following), (II) It depends
only on extra coordinates, and  (III) It preserves the gauge group related to the subgroup of the isometry
group of the bulk space  \cite{confinement}. 
Using the  confining potential $\mathcal{V}$    the matter
fields are exactly localized on the brane and one obtains
\begin{equation}
\alpha_{*}S_{\mu\nu}=8\pi GT_{\mu\nu},~~S_{\mu a}=0,~~S_{ab}=0,
\end{equation}
where $\mu, \nu=0,..., 3$ and $a,b=4,...,n-1$ labels the number of four dimensional
brane and bulk extra dimensions, respectively and
$T_{\mu\nu}$ is the confined matter source on the four dimensional brane.
This is the so called   ``confinement hypothesis''.

Now, it is worth to have a
brief discussion on the bulk and brane energy scales and their corresponding
gravitational coupling constants. The bulk space gravitational coupling constant $\alpha_{*}$ is
\begin{equation}
\alpha_{*} =8\pi G_{n}=\frac{8\pi}{M_{n}^{n-2}},
\end{equation} 
where $G_n$ is equivalently known as the bulk gravitational constant and $M_n$ is the fundamental
energy scale of the bulk space. In a usual four dimensional spacetime, we have
$G_{4}=G= M_{Pl}^{-2}$ where $G$ is the Newtonian gravitational constant. In the static weak field limit of the Einstein field equations, one obtains  $n$-dimensional Poisson equation for the gravitational potential which admits  the following solution
\begin{equation}
V(r)\sim\frac{\alpha_{*}}{r^{n-3}},
\end{equation}
where by supposing that the length scale of the extra dimensions is denoted by $L$, this potential behaves as $V(r)\sim r^{3-n}$,   on scales with 
size $r\lesssim L$, and depends on the number of dimensions of spacetime $n$. On the other hand, for the scales larger than $L$, the 
potential $V(r)$ behaves   as $ V(r)\sim L^{4-n}r^{-1}$ \cite{Maartens}. For $n=4$, we recover the Newtonian four dimensional gravitational potential $ V(r)\sim r^{-1}$.  This means that the Newtonian gravitational constant $G$ or usual Planck scale $M_{Pl}$ are 
effective coupling constants, which describe gravity on the scales much larger than the length scale of extra dimensions, and are proportional  to the bulk fundamental energy scale $M_n$  via \begin{equation}
M_{Pl}^2 =M_{n}^{n-2}L^{n-4},
\end{equation}
where $L^{n-4}$ denotes the volume of the extra dimensional space. This relation indicates that for the extra dimensional volume which is about the Planck scale, i.e. $L\sim M_{Pl}^{-1}$,  we have $M_{n}\sim M_{Pl} $. But for the volume which is significantly above the Planck scale,
we
find  that the  fundamental energy scale of the bulk space $M_n$  is much smaller than the four dimensional effective energy scale $M_{Pl}\sim10^{19}$ GeV.

In order to obtain the effective Einstein field equation induced on the brane,
we consider the following geometrical setup.
 Suppose that the $4D$  background manifold $\mathcal{M}_{4}$
is isometrically embedded
in a $n$ dimensional bulk $\mathcal{M}_{n}$ by a differential map ${\cal Y}^{A}:\mathcal{M}_{4}\longrightarrow\mathcal{M}_{n}$ such that
 \begin{eqnarray}\label{21}
{\cal G} _{AB} {\cal Y}^{A}_{,\mu } {\cal Y}^{B}_{,\nu}=
\bar{g}_{\mu \nu}  , \hspace{.5 cm} {\cal G}_{AB}{\cal
Y}^{A}_{,\mu}\bar{\cal N}^{B}_{a} = 0  ,\hspace{.5 cm}  {\cal
G}_{AB}\bar{\cal N}^{A}_{a}\bar{\cal N}^{B}_{b} = {g}_{ab},
\end{eqnarray}
where ${\cal G}_{AB}$ $(\bar{g}_{\mu \nu })$ is the metric of the bulk
(brane) space $\mathcal{M}_{n}(\mathcal{M}_{4})$,  $\{{\cal Y}^{A}\}$ $(\{x^{\mu }\})$ is the basis
of the bulk (brane), ${\bar{\cal N}^{A}}_{a}$ are $(n-4)$ normal
unit vectors orthogonal to the brane and  $g_{ab}=\epsilon\delta_{ab}$
in which $\epsilon=\pm1$ corresponds to
two possible signatures for each extra dimension of the bulk space.   Perturbation of the background
 $\mathcal{M}_{4}$ manifold in a sufficiently small
neighborhood of the brane along an arbitrary transverse direction $\xi^a$ is
given by
\begin{eqnarray}\label{22}
{\cal Z}^{A}(x^{\mu},\xi^{a}) = {\cal Y}^{A}(x^{\mu}) + ({\cal
L}_{\xi}{\cal Y}(x^{\mu}))^{A}, \label{eq2}
\end{eqnarray}
where ${\cal L}_{\xi}$ represents the Lie derivative along $\xi^a$ denoting the
non-compact extra dimensions. The presence of tangent
component of the vector $\xi$ along the brane can cause some
difficulties because it can induce some undesirable coordinate
gauges. But, it was shown that in the theory of geometric
perturbations, it is quite possible to choose this vector to be
orthogonal to the background manifolds \cite{Nash}. Then, choosing
the extra dimensions $\xi^a$ to be orthogonal to the brane ensures
us about the gauge independency \cite{Jalal} and having
perturbations of the geometrical embedding along the orthogonal extra
directions ${{{\cal N}}}^{A}_{a}$. Thus,  the local coordinates of
the perturbed brane will be \begin{eqnarray}\label{23}
&&{\cal Z}_{,\mu }^{A}(x^{\nu },\xi^a)={\cal Y}_{,\mu }^{A}(x^{\nu })+
\xi^{a}{{{\cal N}}^{A}}_{a,\mu },\nonumber\\
&& {\cal Z}_{,a }^{A}(x^{\nu },\xi^a)={{\cal N}^{A}}_{a}.
\end{eqnarray}
Equation (\ref{22}) implies that since the vectors ${{\cal
N}}^{A}$ depend only on the local coordinates $x^{\mu }$,  ${{\cal
N}}^{A}={{\cal N}}^{A}(x^\mu)$, they will not propagate along the
extra dimensions which can be shown as
\begin{equation}\label{24} {\cal
N}^{A}_{a}={\bar{\cal
N}^{A}}_{\,\,\,\,\,a}+\xi^{b}\left[{\bar{\cal
N}^{A}}_{a},{\bar{\cal N}^{A}}_{b}\right]={\bar{\cal
N}^{A}}_{\,\,\,\,\,a}.
\end{equation}
The above  assumptions lead to the embedding equations of the
perturbed geometry as\begin{eqnarray}\label{25} {\cal G}_{AB}{\cal
Z}_{\,\,\ ,\mu }^{A}{\cal Z}_{\,\,\ ,\nu }^{B}=g_{\mu \nu
},\hspace{0.5cm}{\cal G}_{AB}{\cal Z}_{\,\,\ ,\mu }^{A}{\cal
N}_{\,\,\ a}^{B}=g_{\mu a},\hspace{0.5cm}{\cal G}_{AB}{\cal
N}_{\,\,\
a}^{A}%
{\cal N}_{\,\,\ b}^{B}={g}_{ab},
\end{eqnarray}
where by setting ${{{\cal N}}^{A}}_{a}={\delta^{A}}_{a}$, the
metric of the bulk space ${\cal G}_{AB}$ in the Gaussian frame and
in the vicinity of $\mathcal{M}_{4}$ takes the form of
\begin{eqnarray}\label{26}
{\cal G}_{AB}=\left( \!\!\!%
\begin{array}{cc}
g_{\mu \nu }+A_{\mu c}A_{\,\,\nu }^{c} & A_{\mu a} \\
A_{\nu b} & g_{ab}%
\end{array}%
\!\!\!\right) .
\end{eqnarray}%
Then, the  line element of the bulk space
will have the following form
\begin{equation}\label{27}
dS^{2}={\cal G}_{AB}d{\cal Z}^{A}d{\cal Z}^{B}=g_{\mu \nu
}(x^{\alpha },\xi^a)dx^{\mu }dx^{\nu }+g_{ab}d\xi^{a}d\xi^{b},  \end{equation}
where
\begin{eqnarray}\label{28}
g_{\mu \nu }=\bar{g}_{\mu \nu }-2\xi ^{a}\bar{K}_{\mu \nu a}+\xi
^{a}\xi ^{b}%
\bar{g}^{\alpha \beta }\bar{K}_{\mu \alpha a}\bar{K}_{\nu \beta
b},
\end{eqnarray}%
is the metric of the perturbed brane while $\bar{g}_{\mu\nu}$ is the metric
of original non-perturbed brane (the first fundamental form) and
\begin{eqnarray}\label{29}
\bar{K}_{\mu \nu a}=-{\cal G}_{AB}{\cal Y}_{\,\,\,,\mu }^{A}{\cal
N}_{\,\,\ a;\nu }^{B}=-\frac{1}{2}\frac{\partial
g_{\mu \nu }}{\partial\xi^{a} },
\end{eqnarray}%
is the extrinsic curvature of the original brane (the second
fundamental form). In what follows, we will use the notation $A_{\mu c}=\xi ^{d}A_{\mu
cd}$ where
\begin{equation}\label{210}
A_{\mu cd}={\cal G}_{AB}{\cal N}_{\,\,\ d;\mu }^{A}{\cal N}_{\,\,\
c}^{B}=%
\bar{A}_{\mu cd},
\end{equation}%
which represents the twisting vector fields (the normal fundamental
form). For any fixed extra dimension $\xi^a$, we have  a new
perturbed brane and can define an extrinsic curvature
similar to the original one by
\begin{eqnarray}\label{tildeK}
\tilde{K}_{\mu \nu a}=-{\cal G}_{AB}{\cal Z}_{\,\,\ ,\mu
}^{A}{\cal
N}%
_{\,\,\ a;\nu }^{B}=\bar{K}_{\mu \nu a}-\xi ^{b}\left(
\bar{K}_{\mu
\gamma a}%
\bar{K}_{\,\,\ \nu b}^{\gamma }+A_{\mu ca}A_{\,\,\ b\nu
}^{c}\right).
\end{eqnarray}
Note that the definitions (\ref{26}), (\ref{28}) and (\ref{tildeK}) require
the extrinsic curvature of the perturbed brane to be\begin{eqnarray}
\tilde{K}_{\mu \nu a}=-\frac{1}{2}\frac{\partial {\cal G}_{\mu
\nu
}}{%
\partial \xi ^{a}}.
\end{eqnarray}%
In a geometric setup, the presence of gauge fields $A_{\mu a}$
tilts the embedded family of sub-manifolds with respect to the
normal vector ${\cal N} ^{A}$. According to our construction,
although the original brane is orthogonal to the normal vector
${\cal N}^{A}$, equations (\ref{25}) imply that this will not
true for the deformed geometry. Hence, we change the embedding
coordinates in the following form\begin{eqnarray}\label{212} {\cal
X}_{,\mu }^{A}={\cal Z}_{,\mu }^{A}-g^{ab}{\cal N}_{a}^{A}A_{b\mu
},
\end{eqnarray}%
where the coordinates ${\cal X}^{A}$ describes new family of
embedded manifolds whose members  always will be orthogonal to the
normal vector ${\cal N}^{A}$. In this coordinate system, the
embedding equations of the perturbed brane will be similar to the
original one, described by the relations given in equation (\ref{21}), so
that the coordinates ${\cal Y}^{A}$ are replaced by the new coordinates ${\cal X}^{A}$.
This geometrical  embedding of the new local coordinates will be suitable for
obtaining the induced Einstein field equations on the brane. In this
coordinates, the extrinsic curvature of a perturbed brane is given by
\begin{eqnarray}\label{213}
K_{\mu \nu a}=-{\cal G}_{AB}{\cal X}_{,\mu }^{A}{\cal N}_{a;\nu
}^{B}=\bar{K}%
_{\mu \nu a}-\xi ^{b}\bar{K}_{\mu \gamma a}\bar{K}_{\,\,\nu
b}^{\gamma
}=-\frac{1}{2}\frac{\partial g_{\mu \nu }}{\partial \xi ^{a}},
\end{eqnarray}
which is known as the generalized York's relation and shows the propagation
of the extrinsic curvature because of the propagation of the
metric  in the direction of extra dimensions in the bulk space.
The Gauss-Codazzi equations for the components of the Riemann
tensor of the bulk space in the
embedding vielbein $\{{\cal X}^{A}_{, \alpha}, {\cal N}^A_a \}$ will be
\begin{eqnarray}\label{215}
R_{\alpha \beta \gamma \delta}=2g^{ab}K_{\alpha[ \gamma
a}K_{\delta] \beta b}+{\cal R}_{ABCD}{\cal X} ^{A}_{,\alpha}{\cal
X} ^{B}_{,\beta}{\cal X} ^{C}_{,\gamma} {\cal X}^{D}_{,\delta},
\end{eqnarray}
\begin{eqnarray}\label{216}
K_{\alpha [\gamma c; \delta]}=g^{ab}A_{[\gamma ac}K_{ \delta]
\alpha b}+{\cal R}_{ABCD}{\cal X} ^{A}_{,\alpha} {\cal N}^{B}_{c}
{\cal X} ^{C}_{,\gamma} {\cal X}^{D}_{,\delta},
\end{eqnarray}
where ${\cal R}_{ABCD}$ and $R_{\alpha\beta\gamma\delta}$ are the
Riemann tensors of the bulk and the perturbed brane, respectively \cite{Eisenhart}.
Then, one can find the Ricci tensor by contracting the Gauss equation (\ref{215}) as
\begin{eqnarray}\label{217}
R_{\mu\nu}=(K_{\mu\alpha c}K_{\nu}^{\,\,\,\,\alpha c}-K_{c} K_{\mu
\nu }^{\,\,\,\ c})+{\cal R}_{AB} {\cal X}^{A}_{,\mu} {\cal
X}^{B}_{,\nu}-g^{ab}{\cal R}_{ABCD}{\cal N}^{A}_{a}{\cal
X}^{B}_{,\mu}{\cal X}^{C}_{,\nu}{\cal N}^{D}_{b},
\end{eqnarray}
where a next contraction will give the Ricci scalar as
\begin{equation} \label{218}
R={\cal R}+(K\circ K-K_{a}K^{a})-2g^{ab}{\cal R}_{AB}{\cal
N}_{a}^{A}{\cal N}_{b}^{B}+g^{ad}g^{bc}{\cal R}_{ABCD}{\cal
N}_{a}^{A}{\cal N}_{b}^{B}{\cal N}_{c}^{C}{\cal N}_{d}^{D},
\end{equation}
where use has been made of the notations $K\circ K\equiv K_{a\mu \nu }K^{a\mu \nu }$ and $%
K_{a}\equiv g^{\mu \nu }K_{a\mu \nu }$.
 Consequently, using equations
(\ref{217}) and (\ref{218}), the  relation between Einstein tensors of the bulk and brane   can be obtained as
\begin{equation} \label{219}
G_{AB}{\cal X}_{,\mu }^{A}{\cal X}_{,\nu }^{B}=G_{\mu \nu }+\lambda g_{\mu\nu}-Q_{\mu \nu }-g^{ab}
{\cal R}_{AB}{\cal N}_{a}^{A}{\cal N}_{b}^{B}g_{\mu \nu }+g^{ab}{\cal R}
_{ABCD}{\cal N}_{a}^{A}{\cal X}_{\mu }^{B}{\cal X}_{\nu }^{C}{\cal N}
_{b}^{D},
\end{equation}
where $G_{AB}$, $G_{\mu \nu }$ are the Einstein tensors of the
bulk and brane respectively, and
\begin{equation}\label{220}
Q_{\mu \nu }=\frac{1}{\epsilon}(g^{ab}(K_{a\mu }^{\,\,\,\,\,\,\gamma }K_{\gamma \nu
b}-K_{a}K_{\mu \nu b})-\frac{1}{2}(K\circ K-K_{a}K^{a})g_{\mu \nu }).
\end{equation}
where  $K_a = g^{\mu\nu}K_{a\mu\nu}$ and $K\circ K= K^{a\mu\nu} K_{a\mu\nu}$. From the definition of $%
Q_{\mu \nu }$, it is an independent conserved
geometrical quantity, i.e. $\nabla _{\mu}Q^{\mu \nu }=0$ \cite{Maia1}.

Using the decomposition of the Riemann tensor of the bulk space into the
Weyl curvature tensor, the Ricci tensor and the scalar curvature as
\begin{equation}
\mathcal{R}_{ABCD}=C_{ABCD}-\frac{2}{n-2}\left(\mathcal{G}_{B[D}\mathcal{R}_{C]A}-
\mathcal{G}_{A[D}\mathcal{R}_{C]B}\right)-
\frac{2}{(n-1)(n-2)} \mathcal{R}(\mathcal{G}_{A[D}\mathcal{R}_{C]B}),
\end{equation}
we obtain the induced $4D$ Einstein equation on the brane as
\begin{eqnarray} \label{222}
G_{\mu \nu }&=&G_{AB}{\cal X}_{,\mu }^{A}{\cal X}_{,\nu }^{B}+Q_{\mu \nu }-\mathcal{E}_{\mu\nu}+\frac{n-3}{n-2}g^{ab}
{\cal R}_{AB}{\cal N}_{a}^{A}{\cal N}_{b}^{B}g_{\mu \nu}\nonumber\\
&&- \frac{n-4}{n-2}\mathcal{R}_{AB}{\cal X}_{,\mu }^{A}{\cal
X}_{,\nu }^{B}+\frac{n-4}{(n-1)(n-2)}\mathcal{R}g_{\mu\nu},
\end{eqnarray} where ${\cal E}_{\mu \nu }=g^{ab}{\cal
C}_{ABCD}{\cal X} _{,\mu }^{A}{\cal N}_{a}^{B}{\cal
N}_{b}^{C}{\cal X}_{,\nu }^{D}$ is the electric part of the Weyl
tensor ${\cal C}_{ABCD}$ of the bulk space . From the brane point of
view, the electric part of the Weyl tensor describes a traceless
matter, denoted by dark radiation or Weyl matter. For a
constant curvature bulk space, we have $\mathcal{E}_{\mu\nu}=0$.

Then, the induced Einstein equation, in a constant curvature and
Ricci flat bulk $(\mathcal{E}_{\mu\nu}=\mathcal{R}_{AB}=0)$ will
take the following form
\begin{equation}\label{G}
G_{\mu\nu}=8\pi GT_{\mu\nu}+Q_{\mu\nu},
\end{equation}
where $T_{\mu\nu}$ is the confined matter source on the brane and
$Q_{\mu\nu}$ is a pure geometrical energy-momentum source. We
also assume that the spacetime on the brane is isotropic and
homogeneous and so we have Friedmann-Robertson-Walker (FRW) metric
on the brane,
\begin{equation}\label{ds}
 ds^{2}=-dt^{2}+a^{2}(t)\left(\frac{dr^2}{1-kr^2}+r^{2}d\Omega^2\right),
\end{equation}
where $a(t)$ is the cosmic scale factor, $k=+1, -1$ and $0$
corresponds to the closed, open and flat universes, respectively, and $d\Omega^2=d\theta^2
+ sin^{2}\theta d\phi^2$. The confined
matter source on the brane $T_{\mu\nu}$ can be considered in the
perfect fluid form in a co-moving coordinates as
\begin{equation}\label{T}
 T_{\mu\nu}=(\rho + p)u_{\mu}u_{\nu}+ pg_{\mu\nu},
\end{equation}
where $u_{\alpha}=\delta^{0}_{\alpha}$ is the 4-velocity vector of
the fluid, $\rho$ and $p$ are energy density and isotropic
pressure, respectively. For the metric (\ref{ds}), the components
of the extrinsic curvature tensor can be obtained by using the
Codazzi equation as
\begin{eqnarray}\label{K}
 &&K_{00}=-\frac{1}{\dot a}\frac{d}{dt}\left(\frac{b}{a}\right),\nonumber\\
 &&K_{ij}=\frac{b}{a^2}g_{ij},\ i,j=1,2,3,
\end{eqnarray}
where dot denotes the derivative with respect to the cosmic time $t$,
and $b=b(t)$ is an arbitrary function of time \cite{Maia1,Maia2}.
Then, by defining the parameters $h(t)={\dot b}/{b}$ and $H(t)={\dot
a}/{a}$, the components of $Q_{\mu\nu}$ represented by (\ref{220})
take the form of
\begin{eqnarray}\label{Q1}
&&Q_{00}=\frac{1}{\epsilon}\frac{3b^2}{a^4},\nonumber\\
&&Q_{ij}=-\frac{1}{\epsilon}\frac{b^2}{a^4}\left(\frac{2h}{H}-1\right)g_{ij}.
\end{eqnarray}
Similar to the confined matter field source on the brane $T_{\mu\nu}$, the geometric energy-momentum tensor $Q_{\mu\nu}$ can be identified as
\begin{equation}\label{Q2}
 Q_{\mu\nu}=(\rho_{extr} + p_{extr})u_{\mu}u_{\nu}+ p_{extr}g_{\mu\nu},
\end{equation}
where the $\rho_{extr}$ and  $p_{extr}$ denote the
"extrinsic geometric energy density" and "extrinsic geometric
pressure", respectively (the suffix "$extr$" stands for "extrinsic") \cite{Maia1}. Then, using Eqs. (\ref{Q1}) and
(\ref{Q2}) we obtain
 \begin{eqnarray}\label{rho}
 &&\rho_{extr}=\frac{1}{\epsilon}\frac{3b^2}{a^4},\nonumber\\
 &&p_{extr}=-\frac{1}{\epsilon}\frac{b^2}{a^4}\left(\frac{2h}{H}-1\right).
 \end{eqnarray}
Using Eqs. (\ref{G}), (\ref{Q1}) and (\ref{rho}) and separating
the space and time components we arrive at
\begin{equation}\label{bbc}
\frac{\ddot a}{a}+2\left(\frac{\dot
a}{a}\right)^{2}+2\frac{k}{a^2}=4\pi G(\rho-p)+\frac{1}{\epsilon}
\frac{b^2}{a^4}\frac{1}{\dot a b}\frac{d}{dt}(ab),
\end{equation}
and
\begin{equation}\label{bbc1}
\frac{\ddot a}{a}+2\left(\frac{\dot
a}{a}\right)^{2}+2\frac{k}{a^2}=-\frac{4\pi
G}{3}(\rho+3p)+\frac{1}{\epsilon} \frac{b^2}{a^2}\frac{1}{\dot a
b}\frac{d}{dt}\left(\frac{b}{a}\right).
\end{equation}
Eliminating the $\ddot a$ terms gives the following modified
Friedmann equation on the brane
\begin{equation}\label{F}
\left(\frac{\dot a}{a}\right)^{2}+\frac{k}{a^2}=\frac{8\pi
G}{3}\rho+\frac{1}{\epsilon} \frac{b^2}{a^4},
\end{equation}
which possesses a modified term arising from the extrinsic
geometry of the brane in the bulk space. In the next sections, we will study
this braneworld modification to the Friedmann equations with more details.
\section{The Brane Model with Junction Conditions}
Using the Israeal-Darmois-Lanczos junction condition which exactly
provides  the $Z_2$ symmetry\footnote{The $Z_2$ symmetry means that when you approach the brane from one
side and go through it in the bulk, you face with  the same bulk   having reversed normal unit vector to the brane, i.e $\mathcal{N}^{a}\rightarrow -\mathcal{N}^a$. Indeed, in the presence of  $Z_2$ symmetry, the original non-perturbed  brane located at $\xi^a = 0$ acts
as a mirror for all objects that feels the extra dimensions. The $Z_2$ symmetry
governs for any
perturbation of the original brane leading to a mirror
perturbation  on the other side of the brane \cite{Nash}.}(mirror symmetry)
\cite{Israel}
(see
\cite{Maia1} for a brief review), one can obtain the extrinsic
curvature tensor component of the original non-perturbed brane in
terms of the confined matter sources on brane as
$k_{11}=b(t)=-\alpha_{*}^{2}\rho a^{2}$ \cite{Maia1}.   Then,  the
modified Friedmann equations (\ref{bbc}), (\ref{bbc1}) and (\ref{F}) will take the forms
\begin{equation}
\left(\frac{\dot a}{a}\right)^{2}+\frac{k}{a^2}=\frac{8\pi
G}{3}\rho+ \frac{1}{\epsilon}\alpha_{*}^{4}\rho^{2},
\end{equation}
and
\begin{equation}\label{nn}
\frac{\ddot a}{a}=-\frac{4\pi G}{3}(\rho + 3p) +\frac{1}{\epsilon}\alpha_{*}^{4}\rho^2,
\end{equation}
which shows the $\rho^2$ dependent cosmology \cite{Maia3}.

Now, we intend to obtain modification of the basic law governing the
emergence of space due to the
difference between the  degrees of freedom in the framework of this model.
Using relation ${\ddot a}/{a}=\dot H + H^2 $,  Eq. (\ref{nn}), can be
written as
\begin{equation}\label{HH}
\dot H + H^2=-\frac{4\pi G}{3}(\rho + 3p)
+\frac{1}{\epsilon}\alpha^4_*\rho^2.
\end{equation}
 Multiplying Eq.
(\ref{HH}) by $-4\pi H^{-4}$, we get
 \begin{equation}\label{a}
-4\pi\frac{\dot H}{H^4}=\frac{4\pi}{H^2}
+\frac{4\pi G}{3}\frac{(\rho+3p)}{H^4}4\pi-4\pi\frac{\alpha^4_*\rho^2}{\epsilon
H^{4}}.
  \end{equation}
Assuming $V=4 \pi H^{-3}/3$ as the volume of the sphere on the brane
with Hubble radius $H^{-1}$, we have
\begin{equation}\label{dVt1}
\frac{dV}{dt}=-4\pi\frac{\dot H}{H^4}=\frac{4\pi}{H^2}
+\frac{16\pi^{2}G}{3}\frac{(\rho+3p)}{H^4}-4\pi\frac{\alpha^4_*\rho^2}{\epsilon
H^{4}}.
 \end{equation}
On the other hand, according to Padmanabhan's idea, the number of degrees of
freedom on the spherical surface of Hubble radius $H^{-1}$ is
given by \cite{Pad1}
\begin{equation}
N_{\mathrm{sur}}=\frac{A}{L_{p}^{2}}=\frac{4\pi
}{L_{p}^{2}H^{2}}, \label{Nsur}
\end{equation}
where $L_{p}$ is the Planck length and $A=4\pi H^{-2}$ represents the
area of the Hubble horizon. Using  the
area law $S=A/4{L_{p}^{2}}$, as the saturation of Bekenstein limit \cite{Bek},
we can write\footnote{The Bekenstein limit is an upper limit on the entropy or information that can be contained within a given finite region of space which has a finite amount of energy. It implies that the information necessary to perfectly describe a system, must be finite if the region of space and the energy is finite.\textcolor[rgb]{1,0,0.501961}{\\
}}
\begin{equation}
N_{\mathrm{sur}}=4S. \label{Nsur}
\end{equation}
 Also, the bulk degrees of freedom obey the equipartition law
of energy
\begin{equation}\label{Nbulk}
N_{\mathrm{bulk}}=\frac{2|E|}{k_{B}T}.
\end{equation}
where $E$, $k_B$ and $T$ are the energy inside of the bulk, the Boltzmann constant and the temperature of the bulk, respectively. 
 In the following, we use the units of $k_{B}=c=\hbar=G=L_{p}=1 $ for simplicity.
We also assume the temperature associated with the Hubble horizon
as the Hawking
temperature $T=H/2\pi$, and the energy contained inside  the Hubble volume
in Planck units $V=4\pi /3H^{3}$  as the Komar energy
\begin{equation}\label{Komar}
E_{\mathrm{Komar}}=|(\rho +3p)|V.
\end{equation}

The novel idea of Padmanabhan is that the cosmic expansion,
conceptually equivalent to the emergence of space, is being driven
towards holographic equipartition, and the basic law governing the
emergence of space must relate the emergence of space to the
difference between the number of degrees of freedom in the
holographic surface and the ones in the emerged bulk \cite{Pad1}.
Using equations (\ref{Nbulk}) and (\ref{Komar}), the bulk degrees of freedom may be obtained as
\begin{equation}
N_{bulk}=-\frac{16\pi^2}{3}\frac{(\rho+3p)}{H^4},
\label{Nbulk1}
\end{equation}
where it is assumed that $\rho+3p<0$. Thus, Eq. (\ref{dVt1}) can be
written as  
\begin{equation}\label{dVt}
 \frac{dV}{dt}=N_{sur}- N_{bulk} - N_{extr},
 \end{equation}
where
\begin{equation}\label{N1}
 N_{extr}=\left(\frac{3}{2\pi\epsilon}\right)
\frac{\alpha^4_{*}\rho^2V}{{T}},
 \end{equation}
appears as the number of degrees of freedom corresponding to the extrinsic
geometry of the embedded brane in a higher dimensional spacetime. Indeed,
there are three modes of degrees of freedom, the surface degrees of freedom,
the bulk degrees of freedom and the ones that are related to the extrinsic
geometry of the embedded brane. Since $N_{extr}$ represents the number of degrees
of freedom, it must be positive.
Equation (\ref{N1}) shows that the positiveness of $N_{extr}$ demands for $\epsilon=+1,$  representing
a spacelike extra dimension. Therefore,  it turns out that by applying the Israeal-Darmois-Lanczos junction condition, the timelike extra dimension will be ruled out. This indicates that unlike the other braneworld scenarios
where there is no essential requirement for $\epsilon$ being positive, in
the present scenario the positiveness of $\epsilon$ is imposed by the equation
(\ref{N1}). Generally, the braneworlds with different extrinsic
geometries have different cosmological evolutions.  It is seen that in this scheme, by applying the Israeal-Darmois-Lanczos junction condition,  the
number of degrees of freedom depends on  the bulk space
energy scale $\alpha_{*}$, the signature of the extra dimensions
 $\epsilon=+1$ and the confined matter density $\rho$ as well as the volume
 $V$ and horizon temperature $T$.

At the end of this section, we remark  that
the presence of  the quadratic energy density in the Friedmann equations which was
initially anticipated as a possible solution to the observed accelerated expansion of the Universe, was
shown to be incompatible with the big-bang nucleosynthesis
\cite{nucleo}. Also,  it is shown that this quadratic $\rho$ term can constrain
the high energy inflationary
regimes in comparison with the observational SDSS/2DF/WMAP data \cite{data}. 
In order to reconcile the above mentioned braneworld  scenario with $Z_2$ symmetry or Israeal-Darmois-Lanczos condition with the observational data,  one may
propose that this scenario should be modified, see \cite{modified}.
Therefore, in the following section we will study the case of a general braneworld embedding
procedure without any simplifying junction condition or  $Z_2$ symmetry.
\section{The General Braneworld Model without any Specific Junction Condition }

We consider the geometric quantity (\ref{Q2}) with the barotropic
equation of state
\begin{equation}\label{wextr}
p_{extr}=\omega_{extr}\rho_{extr},
\end{equation}
where $\omega_{extr}$ is the geometric equation of state
parameter and generally can be a function of time. Using equations
(\ref{rho}) and (\ref{wextr}), we obtain the following equation
for $b(t)$
\begin{equation}\label{dotb}
\frac{\dot b}{b}=\frac{1}{2}\left(1-3\omega_{extr}\right)\frac{\dot a}{a},
\end{equation}
where $\omega_{extr}$ is an unknown function. In general,
solving the above equation is impossible unless the functional form
of $\omega_{extr}$ is given. Let us consider the simple case
where $\omega_{extr}=constant$. In this case Eq.
(\ref{dotb}) can be solved immediately as
\begin{equation}\label{b}
b=b_{0}\left(\frac{a}{a_0}\right)^{\frac{1}{2}(1-3\omega_{extr})},
\end{equation}
where $a_0=a(t_{0})$ is the scale factor of the  Universe at the
present time and $b_0$ is an integration constant representing the
curvature warp of the Universe at the present time. Substituting
the solution (\ref{b}) into equations (\ref{Q1}) gives the components of the
geometric
quantity  in terms of $b_0$, $a_0$ and $a(t)$ as
\begin{eqnarray}\label{211}
&&Q_{00}=\frac{1}{\epsilon}\frac{3b_{0}^{2}}{a_{0}^{1-3\omega_{extr}}}a^{-3(1+\omega_{extr})},\nonumber\\
&&Q_{ij}=\frac{1}{\epsilon}\omega_{extr}\frac{3b_{0}^{2}}{a_{0}^{1-3\omega_{extr}}}a^{-3(1+\omega_{extr})}g_{ij},
\end{eqnarray}
and consequently using equations (\ref{rho})
we get
\begin{eqnarray}\label{rhot}
&&\rho_{extr}=\frac{1}{\epsilon}\frac{3b_{0}^{2}}{a_{0}^{1-\omega_{extr}}}a^{-3(1+\omega_{extr})},\nonumber\\
&&p_{extr}=\frac{1}{\epsilon}\omega_{extr}\frac{3b_{0}^{2}}{a_{0}^{1-3\omega_{extr}}}a^{-3(1+\omega_{extr})}.
\end{eqnarray}
Then, using equations (\ref{rhot}) and (\ref{Q2}), the induced
Einstein equation on the brane (\ref{G}) gives us the following
equation for the confined energy density
\begin{equation}\label{rhot1}
\rho=3\left(\frac{\dot
a}{a}\right)^{2}+\frac{3k}{a^2}-\frac{1}{\epsilon}\frac{3b_{0}^2}{a_{0}^{1-3\omega_{extr}}}a^{-3(1+\omega_{extr})}.
\end{equation}

Note that we have not included the cosmological constant   because it is possible to construct a
geometrical origin for the dark energy in a general geometrical embedding
scheme with a brane possessing an extrinsic curvature, to recover
the acceleration of the Universe \cite{Maia1, Jalalzadeh}. The
generalization to the case that the cosmological constant is not
zero, is trivial. Similarly, the confined isotropic pressure
component can be obtained from equations (\ref{G}), (\ref{Q2}) and
(\ref{rhot}) as
\begin{equation}\label{pt1}
p=-2\frac{\ddot a}{a}-\left(\frac{\dot a}{a}\right)^{2}-
\frac{k}{a^2}-\frac{1}{\epsilon}\frac{3b_{0}^{2}\omega_{extr}}{a_{0}^{1-3\omega_{extr}}}a^{-3\left(1+\omega_{extr}\right)}.
\end{equation}
Combining these equations leads to the following equation
\begin{equation}\label{ddota3}
\frac{\ddot a}{a}=-\frac{4\pi}{3}(\rho +
3p)-\frac{1}{3\epsilon}\frac{1}{ a^4}
b^{2}_{0}(1+3\omega_{extr})\left(\frac{a}{a_0}\right)^{1-3\omega_{extr}}.
\end{equation}
Using the same procedure as in the previous section, we obtain
\begin{equation}\label{dvt3}
 \frac{dV}{dt}=N_{sur}- N_{bulk}- N_{extr},
\end{equation}
where the number of degrees of freedom related to the extrinsic
geometry of spacetime has the general form \begin{eqnarray}
N_{extr}&=&4\pi H^{-3}\frac{\frac{1}{3\epsilon}\frac{1}{ a^4}
b^{2}_{0}(1+3\omega_{extr})\left(\frac{a}{a_0}\right)^{1-3\omega_{extr}}}{2\pi\frac{H}{2\pi}}\nonumber\\
&=&\frac{ V}{2\pi T}\frac{1}{\epsilon a^4}
b^{2}_{0}(1+3\omega_{extr})\left(\frac{a}{a_0}\right)^{1-3\omega_{extr}}.
\end{eqnarray}

For the case of a general geometric embedding scheme, the number
of degrees of freedom related to the geometric embedding state of
the brane in a higher dimensional bulk beside the scale factor of the Universe
$a$, depend on the signature of
extra dimensions $\epsilon$, the volume $V$ and
horizon temperature $T$ and warp
factor of the Universe $b_0$ as well as the equation of state parameter
of the geometric fluid $\omega_{extr}$.
If the curvature warp of the
universe $b_0 $ vanishes,  all of the extrinsic curvature components will also
vanish, and the braneworld will behave just as a trivial
plane. In this case, the  number of degrees of freedom corresponding to 
the extrinsic curvature  vanishes and we recover the original Padmanabhan's
relation $\frac{dV}{dt}=N_{sur}- N_{bulk}$.  

Another interesting result is that when the geometric equation
of state parameter becomes $\omega_{extr}=-1/3$, the corresponding degrees of freedom also vanishes. Moreover, there are two possibilities for satisfying
the positivity of $N_{extr}$. The first possibility is      $\omega_{extr}>-1/3$ with $\epsilon=+1$
which indicates a spacelike extra dimension. For this case,  all of the known energy conditions such as weak, null, strong
and dominant energy conditions are satisfied for the geometric fluid. The
second possibility is $\omega_{extr}<-1/3$ with  $\epsilon=-1$,  which accounts
for a timelike extra dimension. In this case, the energy conditions may be violated by the geometric fluid.
It is worth mentioning that, according to (\ref{rhot}),  satisfying the weak
energy condition requires $\epsilon=+1$ which is the same result coming from the positiveness of $N_{extr}$.

Our Universe, is not pure de Sitter but we know that it evolves
toward an asymptotically de Sitter phase. For the purpose of
reaching  holographic equipartition  we need to have
${dV}/{dt}\rightarrow0   $ in the equation (\ref{dvt3}) which
leads to $N_{sur}=N_{bulk}+ N_{extr}$. In order to understand the
prominent feature of $N_{extr}$ it is better to look at
equation (\ref{dvt3}) without this term. Following the discussion of Padmanabhan,
one can consider that $N_{bulk}$  consists of two terms,  one  related to the
normal matter with $\rho+3p>0,$ and the other one related to the dark
energy with $\rho+3p<0$ \cite{Pad1}. So, it is possible to divide the degrees of freedom
of bulk into two terms, one coming from the degrees of
freedom of dark energy leading to acceleration and the
other one coming from the degrees of freedom of normal matter leading to deceleration. Then, equation (\ref{dvt3}) takes the form of $\frac{dV}{dt}= N_{sur}+N_{m}-N_{de}$. Thus, it is seen that a universe without   a dark energy component has no
hope of reaching the holographic equipartition \cite{16}. 

We can remark the important results of the present study as follows.
\begin{itemize}
 \item We can avoid of the  term $N_{de}$, namely dark energy or cosmological constant, which has been proposed by Padmanabhan.  In our general setup, dark energy has a
completely geometrical origin \cite{Maia1}. In fact, the geometrical component denoted by $N_{extr}$ plays the role of $N_{de}$ proposed by Padmanabhan. Similarly, we can understand equation (\ref{dvt3}) in a better
way if we separate out the matter component resulting in deceleration from
the geometrical  component resulting in acceleration. For the sake of simplicity,
we will assume that the Universe has just two components, the normal matter
with $ \rho + 3p>0$
and an effective (geometric) matter   with $\rho_{extr}+3p_{extr}<0$
. By our consideration, equation (\ref{dvt3}) can be expressed in an equivalent
form as
\begin{equation}\label{57}
\frac{dV}{dt}=N_{sur}+N_{m}-N_{extr},
\end{equation}
where $N_{sur}, N_{m}, N_{extr}$ are positive with 
\begin{equation}
(N_{m}-N_{extr})=(\frac{2V}{K_{B}T})[(\rho+3p)+|\rho_{extr}+3p_{extr}|].
\end{equation} 
It is seen that the holographic equipartition condition
with asymptotically vanishing {\it emergence of space} $({dV}/{dt}\rightarrow0)$ can be satisfied only if the Universe possesses $N_{extr}$. Equivalently, the existence of a geometric term (due to
the embedding of the brane) is required for the asymptotic holographic equipartition
which leads the cosmos to find its equilibrium. In the presence of $N_{extr}$, the emergence of space will  lead to $N_{extr}$ dominating over $N_{m}$ with the universe experiencing accelerated expansion due to dark energy. Asymptotically, $N_{extr}$ will approach $N_{sur}$ and  ${dV}/{dt}\rightarrow0$ in a de Sitter universe.

 \item We can keep the term $N_{de}$ as the dark energy degrees of freedom
in the bulk and consider the new term $N_{extr}$ as a geometric contribution
to the term $N_{de}$. In this regard, we can write the following equation
\begin{equation}\label{57}
\frac{dV}{dt}=N_{sur}+N_{m}-(N_{de}+N_{extr}).
\end{equation}
The holographic equipartition condition is satisfied if $N_{sur}+N_{m}=(N_{de}+N_{extr})$.
Moreover, we have
\begin{equation}
[N_{m}-(N_{de}+N_{extr})]=(\frac{2V}{K_{B}T})[(\rho+3p)+|\rho_{de}+3p_{de}|+|\rho_{extr}+3p_{extr}|].
\end{equation} 
In the presence of $(N_{de}+N_{extr})$, the emergence of space will  lead to $(N_{de}+N_{extr})$ dominating over $N_{m}$ with the universe experiencing accelerated expansion due to dark energy and geometric embedding. Asymptotically, $(N_{de}+N_{extr})$ will approach $N_{sur}$ and  ${dV}/{dt}\rightarrow0$ in a de Sitter universe.
\end{itemize}    
\section{Thermodynamics of a General Braneworld Scenario}
The first law of thermodynamics for apparent horizon\footnote{At each hypersurface
of constant time, the apparent horizon of an observer located at
$r=0$ is defined as the sphere whose orthogonal ingoing future-directed
light-rays have vanishing expansion.} reads as \cite{Bousso}
\begin{equation}\label{mmm}
-dE=TdS,
\end{equation}
where $T$ is the time-dependent temperature
of a  thermal heat bath as is perceive by an observer at $r=0$, $dE$ is the change in the mass of
the matter present on the observer's side of the horizon, and $dS$ is the
increases in the horizon entropy.
 In this section, we confirm the validity of the first law of thermodynamics
for apparent horizon in the presence of the additional terms due to the  extrinsic curvature of a  braneworld model and then we study the
second law of thermodynamics with the assumption that in the apparent horizon
the space-time has thermodynamical behaviour. We also show that the second law of thermodynamics for apparent
horizon is always satisfied for an expanding universe.

The entropy of system is obtained as the sum of the surface entropy and the internal entropy. The internal entropy includes
the entropy related to the ordinary matter fields localized on the
brane and the geometric entropy corresponding to the induced
geometric matter.   In this section, we consider a
general embedding scheme without any specific junction condition.

For a flat universe $k=0$, we consider a perfect fluid form for the geometric fluid
$Q_{\mu\nu}$ as \textcolor[rgb]{1,0,0.501961}{in equation (\ref{Q2}) and  using} equations (\ref{rhot}),
(\ref{rhot1}), (\ref{ddota3}) and conservation equation
\begin{equation}
G^{\mu\nu}_{~~;\mu}=(8\pi  T^{\mu\nu}+Q^{\mu\nu})_{;\mu}=0.
\end{equation}
 we obtain
three following equations
\begin{equation}\label{59}
\dot\rho +\dot\rho_{extr}+3H(\rho+p +\rho_{extr}+p_{extr})=0,
\end{equation}
\begin{equation}
\dot H +H^2=-\frac{4\pi }{3}(\rho+3p +\rho_{extr}+3p_{extr}),
\end{equation}
\begin{equation}\label{61}
H^2=\frac{8\pi }{3}\rho+\frac{8\pi }{3}\rho_{extr}.
\end{equation}
As is seen from the above equations, there are two kind of matter sources  for these equations. The first one is the normal matter $\rho$ confined
on the brane and the second one is the induced geometric matter $\rho_{extr}$. In order to obtain the entropy expression associated with
the normal and geometric matter, we consider the variation of their corresponding energies, $dE_{A_m}$ and $dE_{A_{extr}}$,
which are
achievable by the energy crossing formula on the apparent
horizon as \cite{Bousso}
\begin{equation}\label{x}
-dE_{A_m}=4\pi R^{2}T_{\mu\nu}K^{\mu}K^{\nu}dt=4\pi R^{2}(\rho+p)dt,
\end{equation}
\begin{equation}\label{y}
-dE_{A_{extr}}=4\pi R^{2}Q_{\mu\nu}K^{\mu}K^{\nu}dt=4\pi R^{2}(\rho_{extr}+p_{extr})dt,
\end{equation}
where we have used the subscript $A$ to denote for the quantities on the ``Apparent horizon". In order to achieve the total energy crossing formula, we  should consider equations (\ref{x})
and (\ref{y}) together with $dE_{A}=dE_{A_m}+dE_{A_{extr}}$. Then, we obtain
\begin{equation}\label{62}
 -dE_{A}=4\pi R^2(T_{\mu\nu}+Q_{\mu\nu}
)K^{\mu}K^{\nu}dt=4\pi R^2(\rho +p +\rho_{extr}+p_{extr})dt.
\end{equation}
From the equation (\ref{59}) we have
\begin{equation}\label{63}
\rho +p +\rho_{extr}+p_{extr}=-\frac{\dot\rho+\dot\rho_{extr}}{3H}.
\end{equation}
Also, the derivative of equation (\ref{61}) will give 
\begin{equation}\label{64}
2\dot HH=\frac{8\pi }{3}(\dot\rho +\dot\rho_{extr}).
\end{equation}
Then,  using the equations (\ref{63}) and (\ref{64}), we obtain
\begin{equation}\label{65}
\rho +p +\rho_{extr}+p_{extr}=-\frac{2\dot H}{8\pi}.
\end{equation}
Inserting the equation (\ref{65}) into the equation (\ref{62}) and considering $H^{-1}$
instead
of $R_A$, the energy crossing term will take the form of
\begin{equation}\label{A}
dE_{A}=H^{-2}\dot Hdt.
\end{equation}
We note that the radius of apparent horizon is
$R_A=(H^2+\frac{k}{a^2})^{\frac{-1}{2}}$ where for the
flat spatial geometry $k=0$,  the radius of the apparent horizon
$R_A$ coincides with hubble horizon $R_A=\frac{1}{H}$ .
Thus, by taking into account the area law of the
entropy, i.e $S_{A}= {A}/{4}={4\pi H^{-2}}/{4}$, we find
\begin{equation}\label{67}
T_{A}dS_{A}=\frac{H}{2\pi}d\left(\frac{4\pi
H^{-2}}{4}\right)=-H^{-2}\dot Hdt,
\end{equation}
which denotes the total surface crossing entropy of the system.
Thus, one can confirm  the validity of the first law of thermodynamics
using equations (\ref{A}) and (\ref{67}) as
\begin{equation}\label{law}
-dE_A=T_AdS_A,
\end{equation}
where  $T_A={(2\pi
R_A)^{-1}}$. Using
(\ref{67}), we obtain
 \begin{equation}\label{hamed}
 \dot S_A=-2\pi\dot HH^{-3}.
 \end{equation}
On the other
hand, we have the internal entropy $S_I$ for the system which is related to the volume
inside the horizon (we have used the subscript $I$ to denotes for the quantities ``Inside
apparent horizon''). For the internal entropy, we have
\begin{equation}\label{70'}
T_{I}dS_{I}=PdV +dE_{I},
\end{equation}
where $P= p+p_{extr }$ and $E_{I}=( \rho+\rho_{extr})V.$  Variation of
(\ref{70'}) results in 
\begin{equation}\label{s}
\dot S_I=\frac{(\rho+ \rho_{extr} +p +p_{extr})\dot
V+V(\dot\rho+\dot \rho_{extr})}{T_I}=\dot S_{m} + \dot S_{extr},
\end{equation}
where we have divided it to the  entropy corresponding to the normal matter $S_m$ and the geometric matter $S_{extr}$. The extrinsic geometric entropy shows its effect through
the induced geometric fluid on the brane by adding a new  term to the total
internal entropy.
In the above equation, $T_{I}$ is the temperature of the thermal system inside the horizon.
We  consider a thermal system which is bounded
by an apparent horizon that has reached equilibrium with its internal volume. This assumption allows us to put  $T_{I} = T_{A}$ \cite{sheykhi3}.
By putting $V=\frac{4\pi}{3}H^{-3}$
and using equations (\ref{64}) and (\ref{65}) in the equation (\ref{s}), we obtain
\begin{displaymath}\label{70}
\dot S_{m} + \dot S_{extr}=2\pi\left(\frac{\dot
H^2}{H^5}+\frac{\dot H}{H^3}\right).
\end{displaymath}

Then, the total derivative of the entropy by adding  equation (\ref{70})
to  equation (\ref{hamed}), becomes
\begin{equation}
\dot S_t= \dot S_{m} + \dot S_{extr}+\dot S_A
=2\pi\left(\frac{\dot H^2}{H^5}+\frac{\dot
H}{H^3}\right)-2\pi\dot HH^{-3}.
 \end{equation}
According to the second law of thermodynamics, entropy of the
thermodynamical systems can never decrease. So, the derivative of
the entropy with respect to time is always greater than zero, i.e.,
$\dot S_{t}\geqslant0$, so we have
\begin{equation}\label{73}
\dot S_t=2\pi\dot H^2H^{-5}\geqslant0.
 \end{equation}
For the Universe which expands, $H>0,$ the above equation is always true
 representing that the second law of thermodynamics  always holds.
Then, the entropy of the Universe always increases and it  depends on the normal matter $\rho$ and the induced geometric matter $\rho_{extr}$,
and by looking at equation (\ref{61}), we see that it is independent of the
pressure profiles $p$
and $p_{extr}$.  Therefore, in our braneworld model, similar to the Kaluza-Klein model with
$\dot S_t=\frac{21{\pi}^2}{8G_5H^2}\geqslant0$, the entropy
of the Universe on the  apparent horizon always increases \cite{sharif}.

\section{Conclusion}
 In this paper, we have addressed the following question: what is the
dynamical effect of the extrinsic geometrical embedding of an
arbitrary four dimensional brane in a higher dimensional bulk
space, from Padmanabhan's point of view on the emergent Universe?
We have shown that other than the surface degrees of freedom and
the bulk degrees of freedom, there are new degrees of freedom related to the extrinsic geometry of the brane embedded in a higher dimensional bulk space which may play a basic role in the cosmological evolution. Based
on this scenario, we have corrected the Padmanabhan's relation as $\Delta V /\Delta t=N_{\rm sur}-N_{\rm bulk}-N_{\rm extr}$ where $N_{\rm extr}$ accounts
for the new degrees of freedom corresponding to the extrinsic
geometry of the brane. This term has a contribution to the Padmanabhan's
relation such that it plays the role of a dark energy. In this regard, we have separately investigated the braneworld
scenarios with
and without specific junction conditions. Moreover, we have shown that for
the case of braneworlds with Israeal-Darmois-Lanczos junction condition,  the
number of degrees of freedom is determined by the bulk space
energy scale $\alpha_{*}$, the signature of the extra dimensions
 $\epsilon=+1$ and the confined matter density $\rho$ as well as the volume
 $V$ and horizon temperature $T$. In this case, because of positivity of  the number of degrees of freedom, $N_{\rm extr}$, the possibility of having timelike extra dimension is ruled out.   

For the case of a general geometric embedding scheme, the number
of degrees of freedom related to the geometric embedding state of
the brane in a higher dimensional bulk space, beside the scale factor of the Universe
$a$, depends on the signature of
extra dimensions $\epsilon=\pm1$, the volume $V$,
the horizon temperature $T$ and the warp
factor of  Universe $b_0$, as well as the equation of state parameter
of the geometric fluid $\omega_{extr}$.
If the curvature warp of the
Universe $b_0 $ vanishes,  all of the extrinsic curvature components  vanish
too, and the braneworld behaves just as a trivial
plane. In this case, the corresponding number of degrees of freedom also vanishes and we recover the original Padmanabhan's relation $\frac{dV}{dt}=N_{sur}- N_{bulk}$. Also, when the geometric equation
of state parameter will be $\omega_{extr}=-1/3$, the corresponding degrees of freedom also vanishes in this approach. Moreover, the positivity of of
  the
number of degrees of freedom requires    $\omega_{extr} > -1/3$ for a spacelike extra dimension where $\omega_{extr}< -1/3$ for a timelike extra dimension. For the first case, all of
the known energy conditions as weak, null, strong and dominant energy conditions are satisfied by the geometric
fluid while they may be violated by the geometric fluid for the second case.
 Then, we investigated the thermodynamical aspects of this general braneworld model. We confirmed the validity of the first law of thermodynamics on apparent horizon
in the presence of the additional terms due to the  extrinsic curvature of a  braneworld model and then we studied the
second law of thermodynamics with the assumption that on the apparent horizon
the space-time has thermodynamical behaviour. We found that the second law
of thermodynamics is always satisfied for an expanding universe. It should be noticed that the presence of thermal equilibrium fluctuations and quantum fluctuations
can contribute as the entropy correction, and consequently the number of degrees of freedom will be corrected, correspondingly. This work is under our current study
and will be reported in the near future.

\section*{Acknowledgment}
We would like to sincerely thank the anonymous referee whose useful comments
improved the presentation and important results of this paper.
This work has been supported financially by Research Institute for Astronomy and Astrophysics of Maragha (RIAAM) under research project No.1/4165-91.

\end{document}